# ADEPOS: Anomaly Detection based Power Saving for Predictive Maintenance using Edge Computing

**Sumon Kumar Bose**
School of Electrical and Electronic Engineering
Nanyang Technological University
Singapore 639798
Bose0003@e.ntu.edu.sg

**Bapi Kar**
School of Electrical and Electronic Engineering
Nanyang Technological University
Singapore 639798
bapik@ntu.edu.sg

**Mohendra Roy**
School of Electrical and Electronic Engineering
Nanyang Technological University
Singapore 639798
mohendra.roy@ntu.edu.sg

**Pradeep Kumar Gopalakrishnan**
School of Electrical and Electronic Engineering
Nanyang Technological University
Singapore 639798
pradeepgk@ntu.edu.sg

**Arindam Basu**
School of Electrical and Electronic Engineering
Nanyang Technological University
Singapore 639798
arindam.basu@ntu.edu.sg

October 30, 2018

**ABSTRACT**

In Industry 4.0, predictive maintenance (PM) is one of the most important applications pertaining to the Internet of Things (IoT). Machine learning is used to predict the possible failure of a machine before the actual event occurs. However, main challenges in PM are: (a) lack of enough data from failing machines, and (b) paucity of power and bandwidth to transmit sensor data to cloud throughout the lifetime of the machine. Alternatively, edge computing approaches reduce data transmission and consume low energy. In this paper, we propose Anomaly Detection based Power Saving (ADEPOS) scheme using approximate computing through the lifetime of the machine. In the beginning of the machine's life, low accuracy computations are used when machine is healthy. However, on detection of anomalies as time progresses, system is switched to higher accuracy modes. We show using the NASA bearing dataset that using ADEPOS, we need 8.8X less neurons on average and based on post-layout results, the resultant energy savings are 6.4-6.65X.

## 1 Introduction

Machine failures can be very expensive for industries as they result in unplanned downtime and loss of productivity. Maintenance programs typically reduce downtime of machines through routine, scheduled maintenance. Statistical models and reliability data are used [1] to create maintenance schedules that involve periodically replacing machine parts, regardless of their state of health. Although it results in fewer breakdowns, periodic maintenance is not very cost effective. Whenever a non-failing part is replaced during scheduled maintenance, its remaining useful life is wasted [2]. This is a major motivation for industry to move towards Predictive Maintenance (PM). There have been several previous attempts to employ machine learning (ML) techniques in PM such as for aero-engine control system sensor fault detection [3], and fault diagnostics of rotary machine [4].

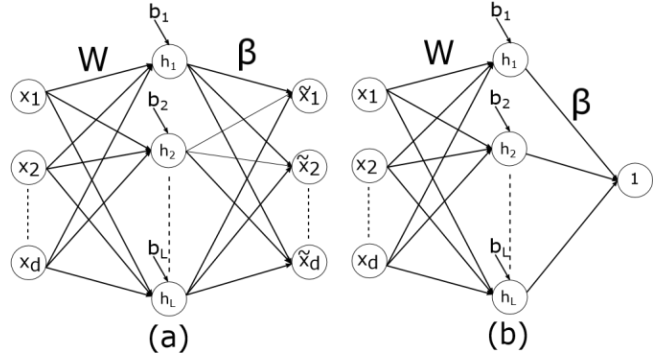

Figure 1: (a) Autoencoder (AE) Architectures: In reconstruction based one class classification (OCC), AE is trained using healthy data to learn input statistics and to reconstruct the input vector at the output. (b) Boundary based OCC is trained to produce a value of 1 at the output for healthy data. During testing when either reconstructed output deviates from input (reconstruction based OCC ) or from 1 (boundary based OCC) by a threshold amount, the machine can be declared as faulty.

Several challenges are associated with Predictive Maintenance systems employing ML techniques that includes nonavailability of adequate amount of data for constructing the prediction model. Adequate information related to machine failures are difficult to obtain [5] as failures are relatively rare and failure signatures vary drastically for different kind of machine's failures. Therefore, learning from a generic dataset may not be an effective approach. It is more reasonable to learn the models from sensor's data attached to the machine. Thus, anomaly detection approach is more suitable for this kind of health monitoring tasks [5], by identifying machine failures based on the deviations from healthy data [6].

Typically, data from IoT sensors are sent to servers for final processing towards failure prediction. This involves large amounts of data transmission and thus lead to significant power consumption and bandwidth requirement. Moreover, unacceptable latencies may be incurred in this approach. Instead, the data processing can be pushed to the edge where bulk of the processing is done near the IoT sensors and only the decision is transmitted to the servers, thus realizing low latency systems. However, this requires the machine learner on edge device to consume low energy to last the lifetime of the machine [7].

Here we present an edge computing framework [7] that exploits anomaly detection for machine health monitoring and reduces energy consumption of the machine learner. We outline the major contributions as: (a) use of ELM-boundary
(ELM-B) based anomaly detector for predictive maintenance with ≈ 20% less convergence time than traditional AE, (b) adaptive use of approximate computing techniques along the age of the machine in order to save energy without compromising on the detection accuracy (ADEPOS), and (c) a hardware architecture to implement the aforementioned approximate computing scheme. While other work have shown hardware architecture for precision scaling of CNN [8] at different steps of the algorithm, our proposal enables even higher power savings by running the entire processor at low precision for most of the life of the machine. The scheme is general and can be used with other approximation enabled processors.

We revisit the preliminaries on autoencoder (AE) and extreme learning machines (ELM) in section 2. Subsequently, we discuss Anomaly Detection based Power Saving (ADEPOS) scheme in section 3. In section 4, we present the results on our experiments on NASA bearing dataset, followed by the concluding remarks.

## 2 Preliminaries

### 2.1 Autoencoder and Extreme Learning Machines

An Autoencoder (AE) is a single layer feed forward neural network, as depicted in Fig. 1(a), consisting of input layer $X = [x_1, x_2, \cdots, x_d]^T$, output layer $\tilde{X} = [\tilde{x}_1, \tilde{x}_2, \cdots, \tilde{x}_d]^T$, and hidden layer $h = [h_1, h_2, \cdots, h_L]^T$, using $d$, $d$ and $L$ neurons respectively. AE is trained to learn input data distribution and to reconstruct input features at its output. The hidden neurons encode an input vector $X$ into feature space $h$ using connection weights $W$ and biases $b = [b_1, b_2, \cdots, b_L]^T$. Likewise, weights ($\beta$) between the hidden and output neurons are used to decode feature space $h$. In boundary based architecture (Fig. 1(b)), the model is trained to produce 1 (or any non-zero real number) at its output [9]. Unknown parameters $W$, $\beta$, $b$ of both architectures are learned using healthy data of the machine since healthy data is readily available. During online monitoring (testing) of machine health when either reconstructed output deviates from input (reconstruction based OCC) or from 1 (boundary based OCC) by a threshold, an anomaly is detected and the machine is declared as faulty.

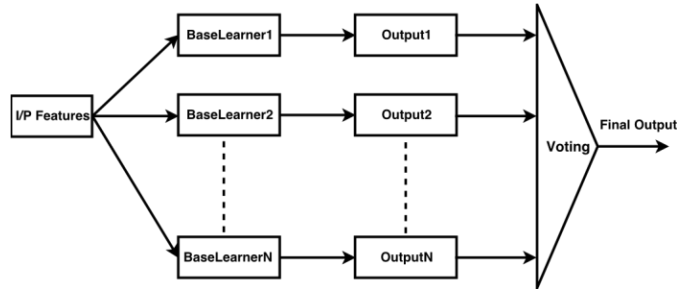

Figure 2: Ensemble Learner containing $N$-base learners (BL) each having $L$ hidden neurons. The final output is decided based on the basis of majority voting of all active BLs.

Details of the encoding and decoding process of AE are presented in Eq. 1 and 2, where $g()$ refers to the neuronal activation function and $h_j$ denotes output of the j-th hidden neuron.

$$h_j = g\left(\sum_{i=1}^{d} W_{ji} x_i + b_j\right); \quad j = 1, 2, ..., L \tag{1}$$

$$\tilde{x}_k = \sum_{j=1}^{L} \beta_{jk} h_j; \quad k = 1, 2, ..., d \tag{2}$$

While g() is typically chosen as *sigmoid*, we found *Rectified Linear Unit* (*ReLU*) to be equally good for AE with the benefit of easier hardware implementation.

Training of traditional autoencoder (TAE) requires a sufficiently large amount of data and iterations for computing optimal $W$ and $\beta$ values. Backpropagation method, although known to yield very accurate models, incurs high computational overhead. An alternative framework, known as *Extreme Learning Machines* (ELM), has been proposed by choosing $W$ and biases $b$ randomly from a continuous probability distribution [10].

In the batch approach [10] for computing least square solution of $\beta$, hidden neuron outputs $H$ are computed for $N$ data samples $\bar{X} = [X_1, X_2, \cdots, X_N]$. Since the desired targets are same as training samples $\bar{X}$ for a reconstruction based AE, the optimal output weight $\beta^*$ is computed as shown in Eqn. 3 where $H^\dagger$ denotes the Moore-Penrose generalized inverse [11].

$$\beta^* = H^\dagger \bar{X} \tag{3}$$

Evidently, the batch approach for computing $\beta^*$ requires a large number of mathematical operations including matrix inversion and multiplication involving $H$ and $\bar{X}$ on a large number of data samples. These computations require large memory and computational energy making them difficult to implement on a sensor node. Alternatively, several online learning techniques such as OSELM and OPIUM have been presented in [12, 13] which have the advantage of not requiring to store the entire training data and reducing computational overhead.

OPIUM method has been adopted for our reconstruction and boundary based one class classification methods due to its smaller computational overhead and reasonable convergence time (see Section 44.1). The equations for updating $\beta$ for each training sample are shown in Eqn. 4, 5 and 6 using initial value of $\theta_{LXL}$ ($\theta_0 = cI$, $c = constant$ and $I = LXL\ Identity\ Matrix$) and an initial estimate of $\beta$ ($\beta_0$) which is computed with a small number of samples ($N_0 \geq L$) [12].

$$\eta_k = \frac{\theta_{k-1} h_k}{1 + h_k^T \theta_{k-1} h_k} \quad (4)$$

$$\beta_k = \beta_{k-1} + \eta_k (X_k - \beta_{k-1} h_k) \quad (5)$$

$$\theta_k = \theta_{k-1} - \eta_k \theta_{k-1} h_k \quad (6)$$

Given the convergence speed of ELM methods and the requirement of lesser output neurons in boundary methods, we combine the advantages and use the ELM boundary method in this work and refer to it as ELM-B in the rest of the paper.

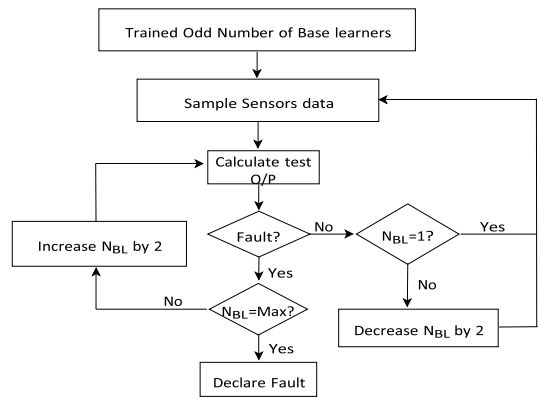

Figure 3: Flowchart for ADEPOS: Approximate computing is achieved by adapting the number of hidden neurons in an Ensemble Learning architecture.

## 2.2 Anomaly Detection and Machine Health Monitoring

In this work, ELM-B based OCC model is trained online during the early phase of machine's life, when the machine is supposed to be in good health. When the learning converges with minimal change in $\beta$, the OCC engine enters into the monitoring (testing) mode. During the early phase of monitoring, the sensor data usually indicates good health of the machine by producing minimal error. If the machine is undergoing any kind of degradation subsequently, the corresponding error from the OCC engine will deviate from the ideal value 1 and can thus be treated as an indication of a potential failure of the machine.

## 3 Our work

In this section, we present an approximate computing strategy–ADEPOS–for machine health monitoring application based on anomaly detection. Though we use ELM-B as the machine learning algorithm, ADEPOS is applicable to other algorithms as well. We also present a VLSI architecture for realizing ADEPOS.

### 3.1 ADEPOS: Anomaly Detection based Power Savings

It is expected that in the beginning of the machine's lifetime, its health condition will have very less degradation. Hence, output of the OCC will be close to ideal and far from the threshold to flag an anomaly. We hypothesize that

<p>we can use very coarse/approximate computing at this initial stage such that the injected errors due to approximation do not cross the anomaly threshold. If an error is observed, we can dynamically increase the accuracy of computation to verify if this is truly an error or a false alarm. Hence, we propose to save power throughout the machine's lifetime based on feedback from the anomaly detector itself justifying the name "ADEPOS". Approximations may be introduced in the network in many ways such as reducing the number of neurons in the network, reducing precision or bit-width of the datapath, reducing accuracy in feature extraction and so on. In this paper, we just show a method to reduce the number of neurons adaptively as a representative example of applying ADEPOS while leaving other approximations for future work.

One of the main challenges of dynamically varying the number of hidden neurons $L$ in the network is that we cannot afford to retrain the new network. Hence, it is best if we can train a larger network in the learning phase and then adaptively shut down parts of this bigger network. However, training a single ELM network with large value of $L$ and then pruning neurons will give incorrect results since the weights obtained during training for neuron 'i' is affected by neuron 'j'. One of the ways to circumvent this problem is to train $N$ different mini ELM-B based models from the input data and create their ensemble [14] as the case with large $L$. We call each mini ELM-B based model as a base learner with $L$ hidden neurons (see Fig. 2).

From a hardware viewpoint, to train a larger neural network of $NL$ neurons we need $N^2L^2$ memory for storing $\theta_k$ (Eq. 4) whereas for $N$ base learners ($L$ neurons for each BL) we need only $NL^2$ memory. This helps to reduce memory area and leakage in the chip implementation. In related work, application of ELM-based adaBoosting method is mentioned in [15] to enhance system resiliency but it has not been exploited for energy optimization.

The flowchart in Fig. 3 shows our proposed algorithm for dynamic increase and decrease of network complexity using ensemble method. At any iteration, effective number of hidden neurons in the network, $L_{eff}$, is given by:

$$L_{eff} = L \times N_{BL} \tag{7}$$

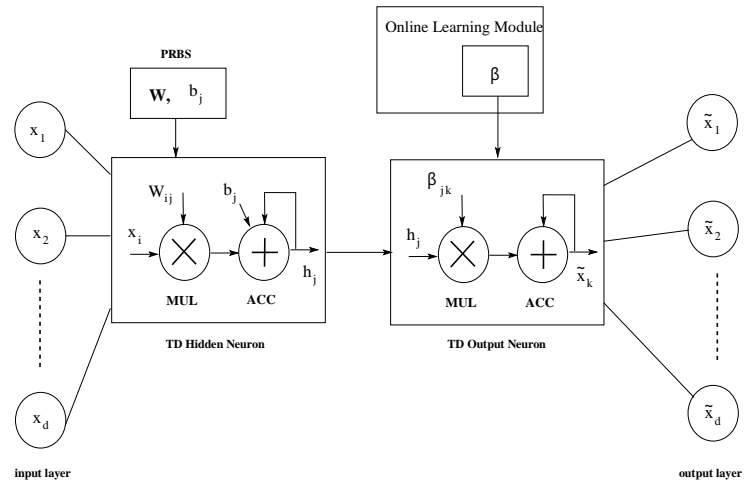

Figure 4: VLSI Architecture of one base learner (ELM based OCC Engine). A system with multiple such base learners can be tuned by ADEPOS to gain energy savings.

where $N_{BL}$ is the number of active or selected base learners (BL) in the network. In the training phase, we trained different BL models to learn the input data statistics. During online testing of the machine health, we start with maximum odd number of BL models available in the network and take the majority voting to decide the output of the anomaly detector. If the system does not raise any alarm for the input sample, we can reduce the number of BL by 2 (since we always need odd number of BLs to get majority voting) at that step. On the other hand, if the system raises an alarm, we immediately increase the number of BLs by 2 to verify if this is truly a fault or just a false alarm. When the anomaly detector raises an alarm and all available BLs are active, we declare a valid fault and call for maintenance. To get power savings, power Supply, $V_{dd}$, of inactive BLs is lowered to the minimum value $V_{dd,retention}$ where its memory can retain its parameters to reduce leakage power.

### 3.2 VLSI Architecture for ELM-AE

Here, we present an overview of the VLSI implementation of the ELM based OCC (Fig. 4). As mentioned earlier, we implemented the online learning framework OPIUM [13] in order to reduce the number of computations per sample as well as to reduce memory area. To reduce logic area, we adopt a minimalist approach with a single neuron each for both hidden and output layers configured to work in *time-division-multiplexing* (TDM) fashion. This architecture also provides the flexibility to configure the system as either a reconstruction or a boundary based ($d = 1$ and $\tilde{x}_1 = 1$) ELM network for OCC (trade-off is higher processing time for reconstruction mode).

The bit width of the data path is configurable from 8 to 16 bits while the number of input and output neurons can vary from 1 to 16. In this design, we use pseudo-random binary sequence (PRBS) module for generating $[W, b]$ for the input layer in order to reduce on-chip memory requirement, while we store the output weights $\beta^*$ in on-chip memory. In order to reduce area and leakage power of the SRAM, we restrict the maximum datapath width to $16 bit$. $V_{dd}$ for the learning module is independent from inference and can be shut off after convergence of online learning. We used TSMC $65 nm$ Low power (LP) technology library for implementing this design. Power analysis was done at $20 MHz$ on the post-layout netlist.

## 4 Results

For validating our work, we use NASA bearing dataset [16] provided by the Center for Intelligent Maintenance Systems
(IMS), University of Cincinnati that is commonly used for testing machine health detection algorithms. Using the information provided in [16], we label the corresponding failed bearings as 1, while 0 is used for the non-failing ones for our cross validation exercise. So far, time, frequency and time–frequency domain analysis of raw bearing data were used extensively for relevant feature extraction [17]. Statistical feature extraction processes on the time-series data were shown to be useful for bearing health monitoring [18]. However, frequency domain based feature extraction techniques were also used for machine health monitoring in the industries to a great extent [19]. In this work, we conduct our experiments on five time-domain features extracted from the raw vibration data contained in NASA dataset, such as (a) RMS, (b) Kurtosis, (c) Peak-Peak, (d) Crest factor, and (e) Skewness since we have validated that this is a minimal set of features that are most informative. Therefore, in our experiments, we set $d = 5$ for obtaining all the parameters, both in MATLAB and hardware simulations.

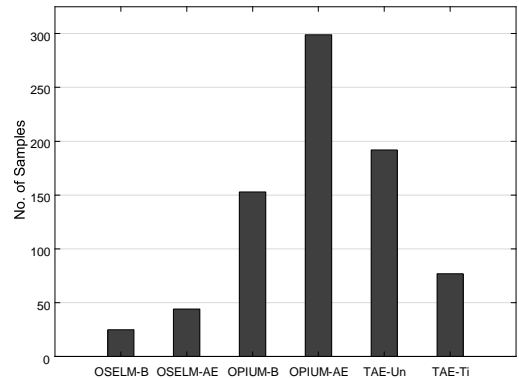

Figure 5: Convergence study of different OCC methods which shows that convergence of OPIUM-B is ≈ 20% faster than TAE-Un.

### 4.1 A comparative study of ELM-based algorithms and traditional AEs

A comparative study of ELM based methods in classifying health conditions of bearings [16] summarized in Fig. 5 shows that ELM based models usually take fewer samples to converge than the traditional AEs . In case of traditional AEs, we use both tied (TAE-Ti) [20] and untied versions (TAE-Un) [21]. For OSELM and OPIUM, we use both boundary and reconstruction based methods. In the above plot, they are represented as OSELM-B, OSELM-AE, OPIUM-B, and OPIUM-AE respectively and the values correspond to the number of training samples taken by the respective models to converge with the same criteria. Though TAE-Ti takes less number of samples, its detection accuracy is poor compared to TAE-Un. Hence, we do not consider it for comparing with ELM-based algorithms. Clearly, ELM-B (OSELM-B/OPIUM-B) and OSELM-AE based OCC methods take fewer samples to converge compared to TAE-Un. Even though OSELM-B converges faster than OPIUM-B, the computational complexity of OSELM-B ($O(L^3)$) is higher than OPIUM-B ($O(L^2)$) as OSELM-B involves matrix inversion during $\beta$ update. Since OPIUM-B takes less computation during training and testing phase, it will consume less power. Hence, we decided to use OPIUM-B algorithm as the online learning algorithm.

### 4.2 ADEPOS: Accuracy vs. Network size

#### 4.2.1 Threshold Selection

As the number of bearing data is limited and there is no information regarding time instance to failure, we utilize a leave-one-out strategy in order to estimate the threshold value ($Thr$) using Eqn. 8. In this threshold calculation method,
we utilized reconstruction errors (difference of reconstructed and expected output) obtained only for good bearings out of all 11 training bearings data. This threshold is then used to test the remaining bearing data. We iterate this approach for each of 12 bearings used as one test dataset.

$$Thr = Max(Err) + 0.5 \times k \times \sigma_{Err} \tag{8}$$

Here, for each good bearing X, we first note the maximum value of testing error, $T_X$ across its lifetime. $Max(Err)$ and $\sigma_{Err}$ are computed as the maximum and standard deviation of these $T_X$ values. In each experimental results presented here, we use $k = 1$. By changing $K$, threshold of the anomaly detector can be varied.

#### 4.2.2 Adaptive Hidden Neurons

In Fig. 6, we present the variation of detection accuracy with respect to effective number of hidden neurons ($L_{eff}$), using ensemble learning method for specific $L$ values. As expected, for same value of $L$, accuracy increases if higher number of base learners ensemble together. To decide on a suitable value of $L$, we generate $N_{BL}$ = (9,7,5) base learners for $L$ = (20,30,40) respectively. It can be seen from the figure that for the same value of $L_{eff}$, creating an ensemble of larger number of base learners with smaller $L$ yields higher accuracy.

Based on the previous study, we fix $L$ = 20 and $N_{BL}$ = 9 and simulate the ADEPOS algorithm described earlier since we obtained almost 100% accuracy at this configuration. Based on the adaptive usage of number of ensembles, we find that the average value of $L_{eff}$ (over 10 trials) throughout the lifetime of all the bearings is only 20.42 without sacrificing the accuracy obtained by using $L_{eff}$ = 180 for $N_{BL}$ = 9 networks in the ensemble. Thus compared to the case of using a fixed value of $L$ = 180 neurons, ADEPOS enables a 8.8X reduction in effective number of neurons.

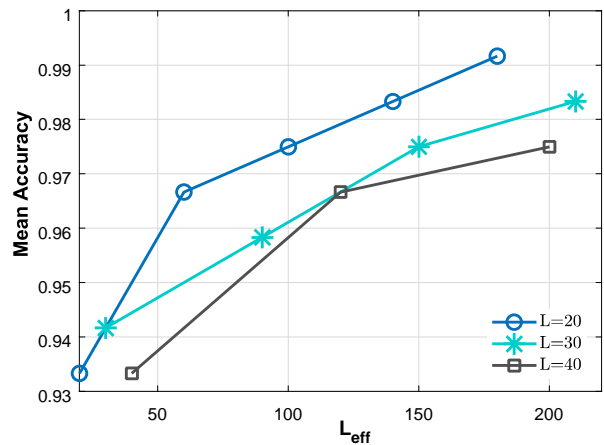

Figure 6: Mean Accuracy vs. number of hidden neurons ($L_{eff}$) for detecting health condition of 12 bearings. Higher number of base learners with small $L$ is preferable.

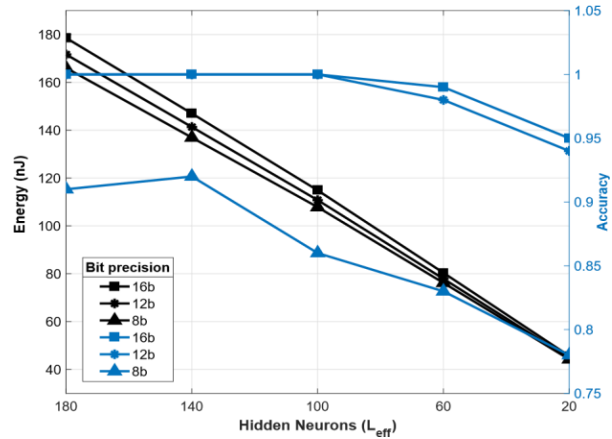

Figure 7: Impact of hidden neurons ($L_{eff}$) and effective bit precision of datapath on Energy and Accuracy

### 4.3   ADEPOS: Energy Savings

In this section, we study the impact of ADEPOS on energy savings through simulation of our VLSI architecture. We also evaluate the effect of bit precision by running simulations with data width of (16,12,8) bits. In power analysis, we varied $L_{eff}$ from 180 down to 20 in steps of 40 and effective bit width from 16$bits$ down to 8$bits$ in steps of 4$bits$. As expected, we notice that the energy consumption reduces from the highest possible accuracy level ($L_{eff}$ = 180 and bit width = 16$bits$), to various lower levels dictated by the values of $L_{eff}$ and the effective bit width of the datapath as shown in Fig. 7. Since detection accuracy degrades drastically at 8$bits$ and lower, we fix the minimum bit precision to 12$bits$. The figure provides an idea of how much energy savings is possible by using ADEPOS to vary the number of hidden neurons using the algorithm depicted in Fig. 3.

Table 1: Energy Consumption of Different algorithms

|  | ELM-AE | ELM-B | ADEPOS |
|---|---|---|---|
| Energy(nJ) | 297.61 | 178.56 | 44.77 |
| Normalized Energy | 6.65 | 3.99 | 1 |

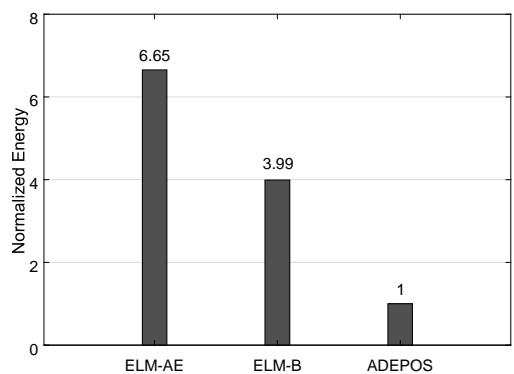

Figure 8: Energy savings by use of ELM-B and ADEPOS mode.

In particular, ADEPOS enables average $6.4 - 6.65X$ reduction in energy for 12-bit and 16-bit datapaths respectively by varying the value of $L_{eff}$. The combined energy savings obtained by choice of algorithm and use of ADEPOS are presented in Fig. 8 and table 1 where 16bit datapath and $L = 180$ are used in ELM-AE and ELM-B.

## 5 Conclusion

In this paper, we present ELM based OCC for anomaly detection in machine health monitoring. Aimed at low energy IoT application for Industry 4.0, our proposed methodology helps to reduce the energy consumption in edge computing devices for anomaly detection. We have shown that ELM based online learning methods require lesser convergence time than TAE. Further, usage of boundary methods as opposed to traditionally used reconstruction methods allow reduction in energy. Further, we propose ADEPOS–a method to save power through approximate computing along the lifetime of the machine by exploiting the fact that it is easy to classify machine health in early part of its life. We adopt ensemble learning method in boundary based OCC that allows us to use a low number of neurons and lower bit width, while dynamically changing to high accuracy mode on detection of anomalies. We show that ADEPOS enables an average energy savings of $6.65X$ when trained to detect failures on NASA bearing dataset.

Though we have only demonstrated ADEPOS by varying the number of neurons and bit width in the network, it is equally applicable to other means of approximation such as reduced sampling rate of raw data. In future, we will explore methods to combine these different approximation modes for a generic solution.

## References

[1] Bryan Dodson. Determining the optimum schedule for preventive maintenance. *Quality Engineering*, 6(4):667–679, 1994.

[2] Seifedine Kadry and Seifedine Kadry. *Diagnostics and Prognostics of Engineering Systems: Methods and Techniques*. IGI Global, Hershey, PA, USA, 1st edition, 2012.

[3] A.N.Vishwanath Rao. Application of auto associative neural network for aero engine control system sensor fault detection, isolation and accomodation. *DRDO Science Spectrum*, pages 12–15, March 2009.

[4] Javier Sanz, Ricardo Perera, and Consuelo Huerta. Fault diagnosis of rotating machinery based on auto-associative neural networks and wavelet transforms. *Journal of Sound and Vibration*, 302(4):981 – 999, 2007.

[5] Shehroz S. Khan and Michael G. Madden. One-class classification: taxonomy of study and review of techniques. *The Knowledge Engineering Review*, 29(3):345–374, 2014.

[6] V. M. Janakiraman and D. Nielsen. Anomaly detection in aviation data using extreme learning machines. In *2016 International Joint Conference on Neural Networks (IJCNN)*, pages 1993–2000, July 2016.

[7] W. Yu, F. Liang, X. He, W. G. Hatcher, C. Lu, J. Lin, and X. Yang. A survey on the edge computing for the internet of things. *IEEE Access*, 6:6900–6919, 2018.

[8] Bert Moons and Marian Verhelst. An Energy-Efficient Precision-Scalable ConvNet Processor in 40-nm CMOS. *IEEE Journal of Solid-State Circuits*, 52(4):903–914, 2017.

[9] Chandan Gautam, Aruna Tiwari, and Qian Leng. On the construction of extreme learning machine for online and offline one-class classification—an expanded toolbox. *Neurocomputing*, 261:126 – 143, 2017. Advances in Extreme Learning Machines (ELM 2015).

[10] Guang-Bin Huang, Qin-Yu Zhu, and Chee-Kheong Siew. Extreme learning machine: Theory and applications. *Neurocomputing*, 70(1):489 – 501, 2006. Neural Networks.

[11] R. Penrose. A generalized inverse for matrices. *Mathematical Proceedings of the Cambridge Philosophical Society*, 51(3):406–413, July 1954.

[12] Nan Ying Liang, Guang Bin Huang, P. Saratchandran, and N. Sundararajan. A fast and accurate online sequential learning algorithm for feedforward networks. *IEEE Transactions on Neural Networks*, 17(6):1411–1423, 2006.

[13] J. Tapson and A. van Schaik. Learning the pseudoinverse solution to network weights. *Neural Networks*, 45:94–100, 2013.

[14] L. K. Hansen and P. Salamon. Neural network ensembles. *IEEE Transactions on Pattern Analysis and Machine Intelligence*, 12(10):993–1001, Oct 1990.

[15] S. H. Wang, H. T. Li, and A. Y. A. Wu. Error-resilient reconfigurable boosting extreme learning machine for ecg telemonitoring systems. In *2018 IEEE International Symposium on Circuits and Systems (ISCAS)*, pages 1–5, May 2018.

[16] Nasa dataset. https://ti.arc.nasa.gov/tech/dash/groups/ pcoe/prognostic-data-repository/.

[17] N Tandon and A Choudhury. A review of vibration and acoustic measurement methods for the detection of defects in rolling element bearings. *Tribology International*, 32(8):469 – 480, 1999.

[18] H.R. Martin and F. Honarvar. Application of statistical moments to bearing failure detection. *Applied Acoustics*, 44(1):67 – 77, 1995.

[19] M. El Hachemi Benbouzid. A review of induction motors signature analysis as a medium for faults detection. *IEEE Transactions on Industrial Electronics*, 47(5):984–993, Oct 2000.

[20] Alain Droniou and Olivier Sigaud. Gated Autoencoders with Tied Input Weights. In *International Conference on Machine Learning*, pages 1–9, United States, 2013.

[21] Sarath Chandar A P, Stanislas Lauly, Hugo Larochelle, Mitesh Khapra, Balaraman Ravindran, Vikas C Raykar, and Amrita Saha. An autoencoder approach to learning bilingual word representations. In Z. Ghahramani, M. Welling, C. Cortes, N. D. Lawrence, and K. Q. Weinberger, editors, *Advances in Neural Information Processing Systems 27*, pages 1853–1861. Curran Associates, Inc., 2014.